\documentclass[a4paper,11pt]{article}
\usepackage{pos}
\usepackage{mathrsfs}
\usepackage{tikz}
\usepackage{fge}
\usepackage{multirow}
\usepackage{psfrag,slashed,cancel,array,graphicx,todonotes,hyperref}
\usepackage{mathtools}
\usepackage{subcaption}
\newcommand{\beq}{\begin{eqnarray}}
\newcommand{\eeq}{\end{eqnarray}}

\newcommand{\rd}{\mathrm{d}}
\newcommand{\rT}{\mathrm{T}}
\newcommand{\abs}[1]{\lvert#1\rvert}
\newcommand{\rs}{\mathrm{s}}
\newcommand{\rt}{\mathrm{t}}

\DeclareMathOperator{\EK}{K}

\title{Topology and geometry of elliptic Feynman amplitudes }
\author{Claude Duhr}
\author*{Yu Jiao Zhu}

\affiliation{\BO}
\emailAdd{cduhr@uni-bonn.de}
\emailAdd{yzhu@uni-bonn.de}

\abstract{
We report on the analytic computation of the 2-loop amplitude for Bhabha scattering in QED.
We study the analytic structure of the amplitude, 
and reveal  its underlying connections to hyperbolic Coxeter groups and arithmetic geometries of elliptic curves.
}

\FullConference{16th International Symposium on Radiative Corrections: Applications of Quantum Field Theory to Phenomenology (
RADCOR2023)\\
28th May - 2nd June, 2023\\
Crieff, Scotland, UK\\}

\begin{document}
\begin{flushright}
BONN-TH-2023-11
\end{flushright}

\def\BO{Bethe Center for Theoretical Physics, Universit\"at Bonn, D-53115, Germany}

\maketitle

  %%%%%%%%%%%%%%%%%%%%%%%%%%%%%%%%%%%%%%%
\section{Introduction}
  %%%%%%%%%%%%%%%%%%%%%%%%%%%%%%%%%%%%%%%
 The first time topology enters into Quantum Field Theory is  when we talk about rotating a fermion. 
 There, 
 the nontrivial element  $-1\in \pi_1(\mathrm{SO}(3;\mathbb{R})\,,\cdot)=\pi_1(\mathbb{RP}^3\,,\cdot)=\mathbb{Z}_2$
 is lifted to the universal covering space $\mathrm{SU}(2)\simeq \mathbb{S}^3$,
 such that any state vector of a fermion is transformed to its antipode $|f\rangle \mapsto e^{i\pi}|f\rangle$.
 The extra phase for a fermion state  under rotation of $2\pi$ has crucial physical consequences, e.g., for the \emph{super selection rules}: 
 it is impossible to prepare states with superpositions of fermions and bosons.
Furthermore, rotation by $4\pi$ is always homotopy to no rotations at all, thus massless particles have helicities of either integers or half integers.

Topology also enters into the dynamics when we take the Fourier-transform of a Green's function 
and take the momenta on-shell.
The result of this procedure is the scattering amplitude, 
whose analytic structure is encoded in the set of symbol letters.
The symbol letters are  locally closed and multi-valued 1-forms over the kinematic base space,
and the general picture is 
to treat the symbol letters as objects of geometric origin, as sheafs of germs of analytic functions~\cite{zbMATH03435823,zbMATH03745550} over the kinematic base space,
which we denote as $(\mathscr{T}(B),q)$.
It turns out that for the phenomenologically relevant processes known to us,  the kinematic base space  $B$ is really special, they are given by
the $n$-dimensional projective space with punctures 
\beq
[s:t:\dots:m^2]\in \mathbb{CP}^n\setminus\Sigma\,,
\eeq
where the punctures $\Sigma$ are the kinematic branch points, which, as we will see, in the case of Bhabha scattering, are given by a union of \emph{linear varieties}.
The covering map $q$ from the covering space $\mathscr{T}(B)$ to  base space $B$ turns out to be special too:
 it is in general normal (Galois)~\cite{zbMATH05832459}, thus the \emph{deck transformation} of $(\mathscr{T}(B),q)$ 
 should have correspondence  to  the automorphisms of Galois field extensions for the meromorphic functions~\cite{zbMATH03745550}.
Besides, the deck group acts transitively on the fibers for normal coverings, and is isomorphic to the \emph{Monodromy }
\beq
\mathrm{Deck}(\mathscr{T}(B),q) \simeq \pi_1(B, \cdot)/ q_* \pi_1(\mathscr{T}(B),\cdot)\simeq\textrm{Mono}\,,
\eeq
which means that the effect of analytic continuation could be depicted globally, without the need to choose a base point. 

In this note, we study several physical  processes--Bhabha scattering and planar top quark production,
we will show several sectors of  the two physics processes  are related to the same moduli space $\mathscr{M}_{1;2}[4]$--the moduli space of elliptic curves with level-4 structure and with one extra marked point,
thus they are partially described  by the same function spaces.

  %%%%%%%%%%%%%%%%%%%%%%%%%%%%%%%%%%%%%%%
\section{The symbol letters}
  %%%%%%%%%%%%%%%%%%%%%%%%%%%%%%%%%%%%%%%
The symbol letters are the closed 1-forms that appear in the canonical differential equations satisfied by the master integrals~\cite{Henn:2013pwa}. They encode the analytic structures of a Feynman amplitude.
It turns out that for the planar master integrals contributing to Bhabha scattering, 
 the set of 1-forms $\omega_i$ are  `dlog' forms (the differential of  logarithmic functions). Here are four typical representatives of them~\cite{Henn:2013woa,Duhr:2021fhk}
\begin{align}
\omega_1(r_{s})=&\rd\!\log(R_1)\,,\quad \omega_2(r_{t})=\rd\!\log(R_2)\,,
\nonumber\\
\quad \omega_3(r_{u})=&\rd\!\log(R_3)\,,\quad \omega_4(r_{st})=\rd\!\log(R_4)\,,
\end{align}
where the $R_i$ are the \emph{symbol letters}, taken from the  \emph {alphabet}
\begin{align}
 \label{eq:dlogs}
\bigg\{
\frac{s-r_s}{s+r_s}\,,
\quad \frac{t-r_t}{t+r_t}\,,
\quad \frac{u-r_u}{u+r_u}\,,
\quad\frac{r_{st}-m^2 s+s t}{r_{st}+m^2 s-s t} 
\bigg\}\,,
\end{align}
and 
 \begin{align}
 \label{eq:sqrts}
r_s  &=  \sqrt{-s} \sqrt{4 m^2-s}\,,
\quad
 r_{st}  =  \sqrt{-s} \sqrt{4 m^6-s (m^2-t)^2}\,,
\nonumber\\  
 r_t &= \sqrt{-t}\sqrt{4 m^2-t}\,, 
 \quad
r_u = \sqrt{-s-t} \sqrt{4 m^2-s-t}\,.
\end{align}
 The square roots $r_s$ and $r_t$ can be rationalized by a degree-2 ramified covering from $\mathbb{CP}^1$ to $\mathbb{CP}^1$
 \beq
\label{eq:xyvar}
\frac{-s}{m^2} = \frac{(1-x)^2}{x}\textrm{~~and~~}\frac{-t}{m^2} = \frac{(1-y)^2}{y}\,,
\eeq
and the 1-forms are converted to meromorphic 1-forms, e.g.\,,
 \beq
\omega_1(x)=&-\rd\!\log(x)\,,\quad \omega_2(y)=-\rd\!\log(y)\,.
\eeq
Most importantly, the symbol letters involve at most simple poles.
The planar topologies contributing to Bhabha scattering are known~\cite{Henn:2013woa,Duhr:2021fhk}. The functions that appear in the alphabet  are algebraic,
for which the uniformizations are well-understood. However, for the non-planar topology, 
the  period functions and integrals over the period functions show up in the alphabet.

  %%%%%%%%%%%%%%%%%%%%%%%%%%%%%%%%%%%%%%%
\section{Period functions for  a family of elliptic curves}
  %%%%%%%%%%%%%%%%%%%%%%%%%%%%%%%%%%%%%%%
We introduce 3 families of elliptic curves on base spaces of dimension 1 and 2 respectively. 
The first  family of elliptic curves that we are going to study is 
 \beq
 \label{eq:E3-def}
 E[2]: Y^2=X(X-1)(X-\lambda)\,, \lambda \in \mathbb{C}\setminus \{0,1\}\,,
 \eeq
and the corresponding  family  of period mappings $\int_\circlearrowleft dz : H_1(E[2]) \to \mathbb{C}$ are defined by
 \begin{align}
  \label{eq:periods-def-G2}
 \Psi_1\left(\lambda\right)\equiv\int_{0}^{\lambda} \frac{\rd X}{Y}
 =2 \EK(\lambda)\,,\quad
 \Psi_2\left(\lambda\right)\equiv\int_{1}^{\lambda} \frac{\rd X}{Y}
 =2 i\EK(1-\lambda)\,,
  \end{align}
  where $\EK$ is the complete elliptic integral of  first kind,
  \beq
  \EK(\lambda) = \int_0^1\frac{\rd t}{\sqrt{(1-t^2)(1-\lambda t^2)}}\,.
  \eeq
  
 The second  family of elliptic curves  is for 2-loop Bhabha scattering:
  \beq
 \label{eq:E4-def-2}
 E_{\textrm{bhabha}}: Y^2=(X-e_1)(X-e_2)(X-e_3)(X-e_4)\,,
 \eeq
with the four roots  given by 
 \begin{align}
 \label{eq:roots-def-2}
e_1= \frac{s}{m^2}-4\,,\quad e_{2,3}=-\frac{s t \pm2 \sqrt{m^2s\, t (s+t-4m^2)}}{m^2(4m^2-t)}\,,
%\quad
%e_3= -\frac{s t -2 \sqrt{m^2s \,t (s+t-4m^2)}}{m^2(4m^2-t)}\,,
\quad e_4=\frac{s}{m^2}\,.
 \end{align}
 The base space can be inferred from the elliptic moduli space, and the cusps correspond to degenerate curves.
By  equating the roots in all possible ways,
  we found the following  varieties
 \begin{align}
 \label{eq:2-dim-punc}
\Sigma=\mathbf{V}(\langle s t(s-4m^2)(s+t)(s+t-4m^2)\rangle) \,,
 \end{align}
 where the  bracket corresponds to  intersections of the ideals generated  by each linear polynomial.
 The  union of the linear varieties  should be deleted from $\mathbb{C}\mathbb{P}^2$, 
 and  base space is the punctured 2-dimensional projective space  $B=\mathbb{C}\mathbb{P}^2\setminus \Sigma$.
  
One of the key objects in this note is the period function for Bhabha 
 \begin{align}
  \label{eq:periods-def}
 \Psi_{\mathrm{bhabha}}\left(\frac{s}{m^2},\frac{t}{m^2}\right)\equiv 2\int_{e_2}^{e_3} \frac{\rd X}{Y}
 =\frac{4 \EK(\lambda)}{\sqrt{(e_1-e_3)(e_2-e_4)}}\,,
%  \nonumber\\
%  \nonumber\\
%  \Psi_2\left(\frac{s}{m^2},\frac{t}{m^2}\right)\equiv \frac{2/\pi}{\sqrt{(-t/m^2)(4-t/m^2)}}\int_{e_1}^{e_2} \frac{\rd X}{\sqrt{Y}}
% \nonumber\\
% =\frac{1/\pi}{\sqrt{(-t/m^2)(4-t/m^2)}} \frac{4 i K(1-\lambda)}{\sqrt{(e_1-e_3)(e_2-e_4)}}\,,
  \end{align}
  where $\EK(\lambda)$ denotes the complete elliptic integral of the first kind and its argument is the \emph{modular $\lambda$ function}.
The shape of the elliptic curves $ E_\textrm{bhabha}$ is parameterized by 
  \beq
  \label{eq:tau-def}
  \tau\equiv \frac{ \Psi_{2}}{ \Psi_{1}}=\frac{i \EK(1-\lambda)}{\EK(\lambda)}\,.
  \eeq
In the next section we will see $\tau$ lives on the \emph{modular curve} $\Gamma_1(4)\backslash\overline {\mathbb{H}}$.
The modular $\lambda$ function,  defined as the cross-ratio of the four roots, is determined by the shape of  the elliptic curve. It can be expressed in terms of {$\theta$-functions}
  \beq
  \label{eq:lambda-def}
  \lambda(\tau)=  \frac{\theta_2^4(\tau)}{\theta_3^4(\tau)}=\frac{4m^2}{2 m^2+\sqrt{\frac{-m^2 s (s+t-4m^2)}{-t}}}\,,
  \eeq 
  where $\theta_i(z,\tau)$ are the standard Jacobi $\theta$ functions, and we define $\theta_i(\tau):=\theta_i(0,\tau)$.

  The last  family of elliptic curves  is for the 2-loop planar top quark production at sector 79~\cite{Muller:2022gec}
\begin{align}
 \label{eq:Etquark-def}
 E_{\textrm{tquark}}: Y^2=\left(X^2-2\left(\frac{t}{m^2}-2\right)X+\frac{t}{m^2}\left(\frac{t}{m^2}-4\right)\right)\left(X^2+2X+1-4\frac{t}{m^2}-4\frac{(m^2-t)^2}{m^2s}\right)\,,
 \end{align}
  and the corresponding period function is 
 \beq
    \Psi_{\mathrm{tquark}}\left(\frac{s}{m^2},\frac{t}{m^2}\right)=
    \frac{
    4 \EK\left(
    \frac{
 \sqrt{  \frac{ m^4+t(s+t-2m^2)}{m^2s}}
    }{3-\frac{t}{m^2}\left(\frac{t}{m^2}-6\right)+\frac{4(m^2-t)^2}{m^2s}+8 \sqrt{  \frac{ m^4+t(s+t-2m^2)}{m^2 s}} }
    \right)
    }
    {
    \sqrt{
    8\sqrt{ \frac{ m^4+t(s+t-2m^2)}{m^2s}}-\left(\frac{t}{m^2}-6\right)\frac{t}{m^2}+3
  +4\frac{(m^2-t)^2}{m^2s}
  }
  }
  \,.
   \eeq

  The general goal is: (1). to find the proper domain (the moduli space of curves) such that on that domain,
   the period functions are converted to single-valued functions;
  (2). to show $\Psi_{\textrm{bhabha}}$ and $\Psi_{\textrm{tquark}}$ are exactly the same as in eq.~(\ref{eq:period-modular}) when expressed through canonical coordinates on the moduli space.
  %%%%%%%%%%%%%%%%%%%%%%%%%%%%%%%%%%%%%%%
  \section{Uniformizations}
    %%%%%%%%%%%%%%%%%%%%%%%%%%%%%%%%%%%%%%%
 %============================================
  \begin{figure}[th]
	\begin{minipage}[t]{.21\textwidth}
		 \subcaptionbox{\label{fig:Name:a}}{\includegraphics[width=\textwidth]{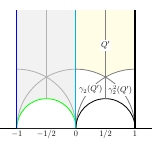}}\\
	\end{minipage}
		\,
	\begin{minipage}[t]{.28\textwidth}
		 \subcaptionbox{\label{fig:Name:b}}{\includegraphics[width=\textwidth]{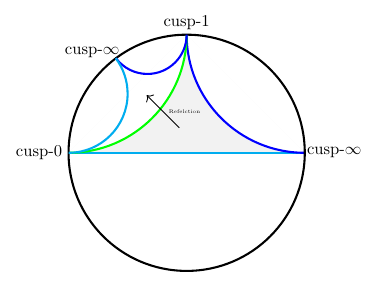}}\\
	\end{minipage}
	\quad
	\begin{minipage}[t]{.19\textwidth}
		\subcaptionbox{\label{fig:Name:c}}{\includegraphics[width=\textwidth]{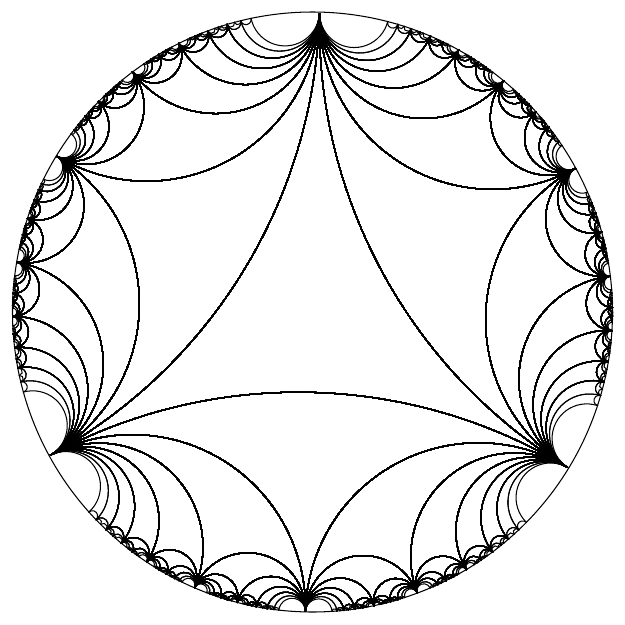}}\\
	\end{minipage}
	\quad
	\,
	\,
	\,
	\,
	\begin{minipage}[t]{.18\textwidth}
		\subcaptionbox{\label{fig:Name:d}}{\includegraphics[width=\textwidth]{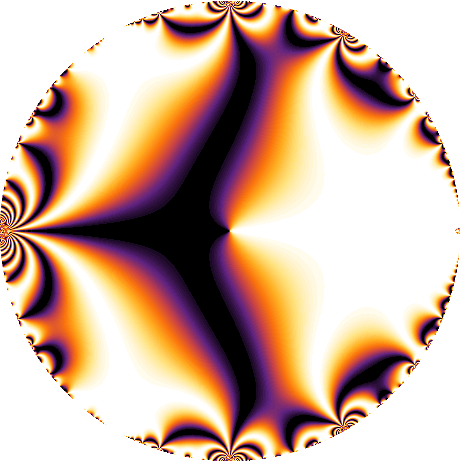}}\\
	\end{minipage}
\caption{The hyperbolic tiling by  triangle group $\Gamma_{\infty\infty\infty}$ and the  modular  function.}
		\label{fig-triangle}
\end{figure}
%============================================    %%%%%%%%%%%%%%%%%%%%%%%%%%%%%%%%%%%%%%% 
    \subsection{Uniformization of punctured $\mathbb{CP}^1$}
   %%%%%%%%%%%%%%%%%%%%%%%%%%%%%%%%%%%%%%%
  By uniformization theorems~\cite{zbMATH00052229}, 
every Riemann surface is the quotient of either $\mathbb{H}\,,$ $\mathbb{C}\,,$ or $\mathbb{C}\mathbb{P}^1$ by a discrete group $\Gamma$ of automorphisms of $\mathbb{H}\,,$ $\mathbb{C}\,,$ or $\mathbb{C}\mathbb{P}^1$. 
Note that, in order for the action of $\Gamma$ to define a covering, $\Gamma$ must act freely, otherwise it is a branched covering.
The universal cover of a (more than twice-) punctured  $\mathbb{C}\mathbb{P}^1$ is $\mathbb{H}$. The reasoning is the following: 
it cannot be $\mathbb{C}\mathbb{P}^1$ since the latter is compact.
Furthermore, the discrete and freely-acting subgroup of Aut($\mathbb{C}$) is a free abelian group with one or two generators. Thus $\mathbb{C}$ is either a covering of a twice punctured Riemann sphere, or a torus.
All remaining Riemann surfaces are essentially isomorphic to $\Gamma\backslash\mathbb{H}$.
   %%%%%%%%%%%%%%%%%%%%%%%%%%%%%%%%%%%%%%% 
    \subsubsection{Poincar\'e polygon theorem}
   %%%%%%%%%%%%%%%%%%%%%%%%%%%%%%%%%%%%%%%
  We start with the uniformization of the thrice-punctured Riemann sphere $\mathbb{CP}^1\setminus \Sigma$ with $\Sigma=\{ 0,1,\infty \}$. Our argument is based on the \emph{Poincar\'e polygon theorem}~\cite{Katok1992,zbMATH06000861,zbMATH05832459},
  the \emph{Riemann mapping theorem} and the \emph{Schwartz reflection principle}.
  Starting from the gray hyperbolic triangle  in fig.~(\ref{fig:Name:a}\ref{fig:Name:b})  with zero angles at the three cusp points,
  we know from the Riemman mapping theorem that the interior of a triangle is conformally equivalent to the upper plane $\mathbb{H}$. We denote such a conformal map by $\lambda(\tau)$. It maps the boundary of the Schwartz triangle to the boundary of the upper half-plane $\mathbb{H}$, which is the real axis with punctures, ${\mathbb{R}}\setminus\Sigma$. 
  By the Schwartz reflection principle, we can perform analytic continuation across the boundaries (see in fig.~(\ref{fig:Name:b})),
  and the image will be the lower plane, because $\lambda(\tau)$ takes real values on the boundaries.
  The reflections can be generated by reflecting  across $z=0\,, \abs{z-1/2}=1/2$ and $z=1$ respectively, and the generators  for the hyperbolic \emph{Todd-Coxeter} group are
  \begin{align}
\label{eq:reflections}
R_1(z)=-\overline{z}\,,\quad R_2(z)=\frac{\overline{z}}{2 \overline{z}-1}\,,\quad R_3(z)=-\overline{z}+2\,.
\end{align}
These generators  are anti-holomorphic M\"obius transformations.
We prefer another set of generators, which are holomorphic:
\begin{align}
\label{eq:generator_G2}
x_1(z)=&R_3\circ R_1(z)=z+2\,,\quad\quad\text{fixes $\infty$}\,,
\nonumber\\
x_2(z)=&R_1\circ R_2(z)=\frac{-z}{2z-1}\,,\quad\quad\text{fixes $0$}\,,
\nonumber\\
x_3(z)=&R_2\circ R_3(z)=\frac{-z+2}{-2z+3}\,,\quad\quad\text{fixes $1$}\,.
\end{align}
Obviously, we have $x_1 x_2 x_3=1$, 
and they are the generators for the Fuchsian triangle group 
 \begin{align}
\Gamma_{\infty,\infty,\infty}\equiv\langle x_1\,, x_2\,,x_3|x_1 x_2 x_3=1\rangle\,,
\end{align}
which are exactly the generators of  the \emph{principal congruence subgroup} $\Gamma(2)$, 
defined as follows
\begin{align}
\Gamma(2)=
\left\{
 \begin{pmatrix}
     a & b \\
    c & d
  \end{pmatrix}
  \in
  \text{PSL}(2,\mathbb{Z})\,\,:\,\,
   \begin{pmatrix}
     a & b \\
    c & d
  \end{pmatrix}
  \equiv\begin{pmatrix}
     1 & 0 \\
    0 & 1
  \end{pmatrix}\,\,\text{mod}\,\, 2
  \right\}\,.
\end{align}
%so we conclude $\Gamma_{\infty\infty\infty}=\Gamma(2)\simeq \pi_1(\hat{\mathbb{C}}\setminus \{0\,,1\,,\infty\} ,\cdot)$.
 Through iterative reflections, on the one hand by the \emph{Poincar\'e polygon theorem}, the hyperbolic triangles will tessellate (fig.~(\ref{fig:Name:c})) the Poincar\'e disk $\mathbb{D}$,
% (if you prefer holomorphic M\"obius transformations, you say the Poincar\'e disk $\mathbb{D}$ is tessellated by the fundamental quadrilateral in fig.~(\ref{fig-triangle})-[b] by iterative modular actions of $\Gamma(2)$), 
and, on the other hand, $\lambda(\tau)$ is analytically continued to the whole Poincar\'e disk. So we conclude, the Poincar\'e disk $\mathbb{D}$ is the universal covering space of thrice-punctured $\mathbb{CP}^1\setminus \Sigma$,
 with $\lambda(\tau)$ the corresponding covering map. Again, from the Schwartz reflection principle, it is easy to see that $\lambda(\gamma\cdot\tau)=\lambda(\tau)\,,\forall\gamma\in\Gamma(2)$, so $\lambda$ descends to a well-defined bijective
 holomorphic map $\tilde\lambda$ from the modular curve $\Gamma(2) \backslash\mathbb{H}$ to $\mathbb{CP}^1\setminus \Sigma$. Its inverse is given by the ratio of the multi-valued period functions.
 %this motivates the next subsection.
  
     %%%%%%%%%%%%%%%%%%%%%%%%%%%%%%%%%%%%%%% 
    \subsubsection{ Torsion data from monodromy group }
   %%%%%%%%%%%%%%%%%%%%%%%%%%%%%%%%%%%%%%%
  Consider the family of elliptic curves $E[2]$ given by eq.~(\ref{eq:E3-def}),
of which  the  corresponding $j$ invariant is
 $ j(\lambda)=256\frac{(1-\lambda(1-\lambda))^3}{\lambda^2(1-\lambda)^2}$. The $j$-invariant
 is ramified at $\lambda=\{0\,,1\,,\infty\}$, each with ramification index $2$, so that deg$j=6$.
This coincides with the index  $[\text{PSL}(2,\mathbb{Z})\,:\,\Gamma(2)]$ of $\Gamma(2)$ in $\text{PSL}(2,\mathbb{Z})$ (see fig.~(\ref{fig:Name:a})).
  This shows that the family of elliptic curves carries extra information other than the shape encoded in $\tau$.
  The extra information turns out to be relevant torsion data for the congruence subgroup $\Gamma(2)$, and $E[2]$ is a family of elliptic curves attached to the moduli space $\mathscr{M}_{1;1}[2]$~\cite{zbMATH02134201}.
  
  The extra torsion data can be uncovered by computing the monodromy group for the corresponding Picard-Fuchs differential equation:
    \begin{align}
  \label{eq:picard-fuchs}
  \left [4 \lambda (1-\lambda) \frac{d^2}{d\lambda^2}+4(1-2 \lambda) \frac{d}{d\lambda}-1\right] \Psi_i=0\,, \quad i=1\,,2.
  \end{align}
  The solution space is $\mathbb{C}\Psi_1(\lambda) \oplus \mathbb{C}\Psi_2(\lambda)$. It is a vector bundle over   $\mathbb{CP}^1\setminus{\Sigma}$.
  By considering analytic continuation with a fixed base point, or equivalently the monodromy action~\cite{zbMATH00747049,zbMATH05832459} by the base space fundamental group $\pi_1(B\,,\cdot)=\mathbb{Z}*\mathbb{Z}$,
  where $B=\mathbb{CP}^1\setminus{\Sigma}$, 
  we have 
   \beq
   \label{mono-rep}
\vec\Psi{\circlearrowleft} (\lambda)= \rho_{[\gamma]}\cdot\vec\Psi (\lambda)\,~~~\textrm{with}~~~\rho_{[\gamma_1][\gamma_2]}=\rho_{[\gamma_1]}\cdot\rho_{[\gamma_2]}\,.
 \eeq
 And the images of $\pi_1(B\,,\cdot)$ under the homeomorphism 
\beq\begin{split}\label{eq:monodromy_rep}
\rho : \pi_1(B\,,\cdot) &\,\to  \textrm{GL}_2(\mathbb{C})\,\\
[\gamma]&\,\mapsto \rho_{[\gamma]}
\end{split}\eeq
turn out to be the two free generators of $\Gamma(2)\lhd\mathbb{P}\text{SL}(2,\mathbb{Z})$
 \begin{align}
\rho_{[{\circlearrowleft}_0]}=
\begin{pmatrix}
     1 & 2 \\
    0 & 1
  \end{pmatrix}\,,
  \quad
  \rho_{[{\circlearrowleft}_1]}=
  \begin{pmatrix}
     1 & 0 \\
    -2 & 1
  \end{pmatrix}\,.
  \end{align}
By the definition of $\tau$ in eq.~(\ref{eq:tau-def}),
we see that analytic continuation induces a modular transformation
\beq\label{eq:Moebiusaction-G2}
\tau_{\circlearrowleft} = \frac{a\tau+b}{c\tau+d} \equiv \gamma\cdot \tau\,,\quad \tau \in \mathbb{H}\,,\quad \gamma\in\Gamma(2)\,.
\eeq
Moreover, 
\beq
\Psi_{1\circlearrowleft}(\lambda)=c \Psi_2(\lambda)+d\Psi_1(\lambda)=(c\tau+d) \Psi_1(\lambda)\,,
\eeq
so the period is a modular form of weight 1. Indeed, by  the pull back of  $\lambda$
 \beq
  \label{eq:lambda-def-2}
 \quad  \Psi_1(\tau)=4\lambda^*(\EK)(\tau)=4\EK(\lambda(\tau))=2\pi\theta_3^2(\tau)\,.
  \eeq 
%  where $q=e^{i\pi\tau}$. 
  This is how period functions are related to modular forms.

     %%%%%%%%%%%%%%%%%%%%%%%%%%%%%%%%%%%%%%%
  \subsubsection{Algebraic realization of the universal family of complex tori $\mathcal{E}_{\Gamma(2)\backslash{\mathbb{H}}}$}
  \label{sec-gamma2}
     %%%%%%%%%%%%%%%%%%%%%%%%%%%%%%%%%%%%%%% 
  As a summary of the previous subsections, we have established the following relations:
\beq
\label{diagram:commu}
\begin{tikzpicture}[scale=2]
    \node (E) at (0,0) {$\mathbb{H}$};
    \node[below=of E] (N) {$\Gamma(2)\backslash\mathbb{H}$};
    \node[right=of N] (M) {$\mathbb{C}\mathbb{P}^1\setminus\{0\,,1\,,\infty\}$};
    \draw[->] (E)--(N) node [midway,left] {$\pi$};
    \draw[->,transform canvas={yshift=+1mm}] (N)--(M) node [midway,above] {$\tilde{\lambda}$};
    \draw[->,transform canvas={yshift=-1mm}] (M)--(N) node [midway,below] {$\Psi_2/\Psi_1$};
    \draw[->] (E)--(M) node [midway,above] {$\lambda$};
\end{tikzpicture}
\eeq
with $\mathrm{Deck}_\lambda (\mathbb{H}\,, B=\mathbb{CP}^1\setminus\Sigma)\simeq\mathrm{Mono}\simeq \pi_1(B\,,\cdot)\simeq\Gamma(2)\simeq\mathbb{Z}*\mathbb{Z}$.
Based on the previous observations, we will  establish a equivalence between 
$\mathcal{E}_{\Gamma(2)\backslash{\mathbb{H}}}\equiv\left(\mathbb{Z}^2\rtimes\Gamma(2)\right)\backslash\left(\mathbb{C}\times{\mathbb{H}}\right)$
and the following family of elliptic curves  
\beq
 E_{\tau}[2]: Y^2=X(X-1)(X-\lambda(\tau))\,, \quad\tau \in \Gamma(2)\backslash \mathbb{H}\,,
 \eeq
 which is conformally equivalent to $E[2]$ given by eq.~(\ref{eq:E3-def}).
To this end, we utilize the  Abel maps from $\mathcal{E}_{\Gamma(2)\backslash{\mathbb{H}}}$ to $ E_{\tau}[2]$,
given by $f_{[2]}$
\beq
(z,\tau) \in \mathcal{E}_{\Gamma(2)\backslash{\mathbb{H}}} \stackrel{f_{[2]}}{\longmapsto} \left[X: \frac{1}{\Psi_1(\tau)}\partial{X}/{\partial z}:1\right] \in  E_{\tau}[2]\,,
\eeq
where after uniformization, 
 the modular weight-1 period $\Psi_1(\tau)$ with respect to $\Gamma(2)$
 is given in eq.~(\ref{eq:lambda-def-2}). The isomorphism  bwtween $\mathcal{E}_{\Gamma(2)\backslash{\mathbb{H}}}$ and $E[2]$  is given here by two equations 
\beq
\label{eq: modular-para-g2}
 X(z,\tau)=\frac{\theta^2_2(\tau)\theta^2_4(z,\tau)}{\theta^2_3(\tau)\theta^2_1(z,\tau)}\,,\quad \lambda(\tau)=  \frac{\theta_2^4(\tau)}{\theta_3^4(\tau)}\,.
\eeq
The function $f_{[2]}$ defined through  eq.~(\ref{eq: modular-para-g2}) is invariant under the action of the semi-direct product $\mathbb{Z}^2\rtimes\Gamma(2)$, 
\beq
f_{[2]}[z\,,\tau]=f_{[2]}[((m,n),\gamma)\cdot(z\,,\tau)]\,,\quad\forall  (m\,,n) \in  \mathbb{Z}^2\,, \gamma\in\Gamma(2)\,,
\eeq
with $((m,n),\gamma)\cdot(z\,,\tau)=\left( \frac{z+m \tau+n}{c \tau+d}\,,\gamma\cdot\tau\right)$.
Thus, $f_{[2]}$  is a well-defined holomorphic map between $\mathcal{E}_{\Gamma(2)\backslash{\mathbb{H}}}$ and $E[2]$.
The map is bijective, its inverse is given by a family of Abel maps, we say $E[2]$ is an algebraic realization of  the universal curve $\mathcal{E}_{\Gamma(2)\backslash{\mathbb{H}}}$.

     %%%%%%%%%%%%%%%%%%%%%%%%%%%%%%%%%%%%%%% 
    \subsection{Uniformization of the punctured $\mathbb{CP}^2$}
   %%%%%%%%%%%%%%%%%%%%%%%%%%%%%%%%%%%%%%%
       %%%%%%%%%%%%%%%%%%%%%%%%%%%%%%%%%%%%%%% 
   \subsubsection{Canonical coordinates on  moduli space $\mathscr{M}_{1;2}[4]$}
     %%%%%%%%%%%%%%%%%%%%%%%%%%%%%%%%%%%%%%%
We want to generalize the story from the previous section to higher dimensions, where the base space is given by $\mathbb{CP}^{2}\setminus\Sigma$, with $\Sigma$ given in eq.~(\ref{eq:2-dim-punc}).
Intuition from lower dimensions tells us that the base should be isomorphic to some moduli space, which encodes the shape $\tau$ of the curves  and the  associated arithmetic data.
Inspired by the  \textrm{Mordell-Weil theorem}, which states that rational points on an elliptic curve  form a finitely-generated  abelian group $T_\text{torsion}\oplus r \mathbb{Z}$, 
 we propose that the marked points should be given by \emph{the generator of Mordell-Weil group}  for the family of elliptic curves  $E_\textrm{bhabha}$.
 We denote such a generator on $E_\textrm{bhabha}$ by $p_0$ with coordinates  given by
\begin{align}
\left[\frac{(s-4 m^2 ) s}{-4m^2 + 2 s +  t} :\, \frac{(s-4 m^2 ) s/m^2 (s+t-4m^2 ) }{( 2 s + t-4m^2)^2/(s + t)} :\, m^2\right]\,.
\end{align}
The corresponding  rational sections generated by $p_0$ are:
\beq
\{ 
[n]p_0| \,p_0 \in T_\text{torsion}\oplus r \mathbb{Z}\,, n\in \mathbb{Z} 
\}\simeq(\mathbb{Z}\,,+)\,.
\eeq
We relabel  $E_\textrm{bhabha}$ by canonical coordinates $(z\,,\tau)$.
The variable $\tau$ indicates the  shape of the curve, with possible \emph{level structures} that encodes the torsion data~\cite{zbMATH02134201}.
The variable $z$ indicates the  infinite subgroup $(\mathbb{Z}\,,+)$ of the \emph{Mordell-Weil group} for a family of elliptic curves.
This  gives us two equations
\begin{equation}\begin{split}
\label{eq:abel_map}
\emph{Abel map:} \quad&\frac{(e_2-e_4)(e_1-X)}{(e_1-e_4)(e_2-X)}=\frac{\theta_2^2(\tau)}{\theta_3^2(\tau)}\frac{\theta_1^2(z,\tau)}{\theta_4^2(z,\tau)}\,, \\
\emph{Modular $\lambda$:} \quad&\frac{4}{2 +\sqrt{\frac{- \rs (\rs+\rt-4)}{-\rt}}}=\frac{\theta_2^4(\tau)}{\theta_3^4(\tau)}\,,
\end{split}\end{equation}
where %$q=e^{i\pi\tau}$ and 
$e_i$ are the four roots of $E_\textrm{bhabha}$ given in eq.~(\ref{eq:roots-def-2}). The $X$-coordinate is $X=\rs(\rs-4)/(-4 + 2 \rs +  \rt)$.
The Mandelstam variables
$\textrm{s}=s/m^2$ and $\textrm{t}=t/m^2$ can be solved from eq.~(\ref{eq:abel_map})  as functions of $(z,\tau)$:
\begin{align}
\label{eq:modular-para-cp2}
\textrm{s}=-\frac{4(-1+R)\times (-2+\lambda)}{-2+\lambda+R \times\lambda}\,,
\quad
\textrm{t}=\frac{4(-1+R)\times R\times\lambda^2 }{({-2+R \times\lambda})({-2+\lambda+R \times\lambda})}\,,
\end{align}
where  $R$ is given by the Abel map and $\lambda$  is the modular $\lambda$ function where
\beq
R=\frac{\theta_2^2(\tau)}{\theta_3^2(\tau)}\frac{\theta_1^2(z,\tau)}{\theta_4^2(z,\tau)}\,,\quad \lambda=\frac{\theta_2^4(\tau)}{\theta_3^4(\tau)}\,.
\eeq
Note that $(\textrm{s}\,, \textrm{t})$ as functions of $(z\,,\tau)$ are invariant under the Hecke subgroup of $\Gamma_0(2)=\Gamma_1(2)> \Gamma_1(4)$.

With the same idea one can reparametrize the family of elliptic curves for top quark production in eq.~(\ref{eq:Etquark-def}). We found, after uniformization, a striking equivalence
\beq
\label{eq:period-modular}
   \Psi_\mathrm{bhabha}\left(z,\tau \right)=\Psi_\mathrm{tquark}(z,\tau)=\frac{\pi \theta_2^2(\tau)}{2}\frac{\theta_3(z,\tau)\theta_4(z,\tau)}{\theta_1(z,\tau)\theta_2(z,\tau)}\,,\,\,
\eeq
which is a modular function of weight 1 with respect to the semidirect product  $\mathbb{Z}^2\rtimes\Gamma_1(4)$,
\beq
\label{eq:modular-g14}
\Psi_\mathrm{bhabha}=\left(\frac{z+m \tau+n}{c \tau +d},\gamma\cdot\tau \right)=\frac{1}{c \tau+d} \Psi_\mathrm{bhabha}\left(z,\tau \right)\,,\quad\forall \gamma \in \Gamma_1(4)\,.
\eeq
Based on these observations, we argue that the base space $\mathbb{C}\mathbb{P}^2\setminus\Sigma$ is biholomorphic to the 
 quotient space of the universal covering space $\mathbb{C}\times\mathbb{H}$ modulo the action of the uniformization group $\mathbb{Z}^2\rtimes\Gamma_1(4)$:
\beq
\label{diagram:commu-high-dimen}
\begin{tikzpicture}[scale=2]
    \node (E) at (0,0) {$\mathbb{C}\times{\mathbb{H}}$};
    \node[below=of E] (N) {$\left(\mathbb{Z}^2\rtimes\Gamma_1(4)\right)\backslash\left(\mathbb{C}\times{\mathbb{H}}\right)$};
    \node[right=of N] (M) {$\mathbb{C}\mathbb{P}^2\setminus\{\text{kinematic branches}\}$};
    \draw[->] (E)--(N) node [midway,left] {$\pi$};
    \draw[->,transform canvas={yshift=+1mm}] (N)--(M) node [midway,above] {$\tilde{f}$};
    \draw[->,transform canvas={yshift=-1mm}] (M)--(N) node [midway,below] {$(z, \Psi_2/\Psi_1)$};
    \draw[->] (E)--(M) node [midway,above] {$f$};
\end{tikzpicture}
\eeq

  \subsubsection{The pullback of the symbol letters for Bhabha}
  \label{pullback-bhabha}
We list several non-trivial closed 1-forms for Bhabha scattering and show their pullbacks.
The first two are the fundamental 1-forms,   
\begin{align}
\label{eq:fundamental_form}
\omega_{\tau}=&
\frac{\rd \rt \,(\rs-4) \rs-\rd \rs\, \rt (2\rs+\rt-4)}{2\rs \rt^2(\rs-4)(\rs+\rt-4)(\rs+\rt )\Psi_1^2(\rs,\rt)}\,,
\nonumber\\
\omega_z=& 
\rd \rt
\,\frac{-1}{4\rt^2(\rs+\rt-4)(\rs+\rt)}\frac{\rT_1(\rs,\rt)}{\Psi_1^2(\rs,\rt)}
\nonumber\\
+&
\rd \rs
\bigg(
\frac{2\rs+\rt-4}{4\rs (\rs-4)  \rt (\rs+\rt )(\rs+\rt-4)}\frac{\rT_1(\rs,\rt)}{\Psi_1^2(\rs,\rt)}
+\frac{2 \sqrt{-\rt}\sqrt{4-\rt}}{\rs (\rs-4) \rt (\rt-4) }\frac{1}{\Psi_1(\rs,\rt)}
\bigg)\,,
\end{align}
where the integral over the period is defined as 
\begin{align}
\label{eq:T12}
\rT_1(\rs,\rt)=&\int
 \rd \rs\,
 \bigg[\frac{-\rt}{\rs}\frac{4\rs^2+4\rs(\rt-4)+\rt(\rt-4)}{ \sqrt{-\rt}\sqrt{4 -\rt}}\Psi_1
-8 \rt\frac{(\rs+\rt-4)(\rs+\rt)}{ \sqrt{-\rt}\sqrt{4-\rt}\,(t+2\rs-4)}\partial_\rs \Psi_1\bigg]
\nonumber\\
 +& \rd \rt
  \bigg[\frac{-\rt}{4-\rt} \frac{-48  +4 \rs+2\rs^2+12\rt+\rs \rt}{ \sqrt{-\rt}\sqrt{4-\rt}\,(\rt+\rs-4)}\Psi_1 \bigg]\,,
\quad \text{~~with~~}\Psi_1=\pi \sqrt{-\rt}\sqrt{4-\rt}\Psi_{\mathrm{bhabha}}\,.
\end{align}

 And a non-trivial depending on the integral $\rT_1(\rs,\rt)$,  
 \begin{align}
\label{eq:modular-example}
\omega_{41}=&\,
\rd \rt\bigg[
\,\frac{1}{2\rt^2(\rs+\rt-4)(\rs+\rt)}\frac{\rT_1^2(\rs,\rt)}{\Psi_1^2(\rs,\rt)}
+\frac{2(\rs-4)}{(\rt-4)(\rs+\rt-4)}\bigg]
+\,\rd \rs
\bigg[
-\frac{2 \rt (2\rs^2+\rs \rt +4  \rs +12   \rt  -48 )}{(\rs-4)\rs (\rt-4)(\rs+\rt-4)}
\nonumber\\
+&\frac{2 \rs+\rt-4}{2 (\rs-4)\rs \rt (\rs+\rt-4)(\rs+\rt)}\frac{\rT_1^2(\rs,\rt)}{\Psi_1^2(\rs,\rt)}
+\frac{\sqrt{\rt(\rt-4)}}{(s-4)\rs(4-\rt)\rt}\frac{\rT_1(\rs,\rt)}{\Psi_1(\rs,\rt)}
\bigg]
\,.
\end{align}
The following results show  the magic effect of the map given in eq.~(\ref{eq:modular-para-cp2}),
\begin{align}
\label{eq:fundamental-differential}
\omega_\tau\xmapsto{~f^*} i\pi \rd\tau \textrm {~~~and~~~} \omega_z\xmapsto{~f^*} 2\pi \rd z\,,
\end{align}
and also the example with non-trivial connections to   Kronecker's differential forms~\cite{Weinzierl:2022eaz,Broedel:2017kkb,zagier1991periods},
\begin{align}
\label{eq:w-41}
\omega_{41}\xmapsto{~f^*} 8 \omega_2^{\text{Kro } }(2 z\,, \tau)-8 \omega_2^{\text{Kro} }(2 z\,, 2\tau)+\frac{4}{3}\frac{\rd q}{q}\Theta_{D_4}(q^2)\,,
\end{align}
where $q=e^{i\pi\tau}$ and the $\theta$-series of the $D_4$ root lattice is given by
\beq
\Theta_{D_4}(q^2)=\theta_3^4(2\tau)+\theta_2^4(2\tau)\in\mathcal{M}_2(\Gamma_0(2))\subset \mathcal{M}_2(\Gamma_1(4))
\,.\,\,\,
\eeq

 \subsubsection{Algebraic realization of Kronecker's differential forms}
In section~(\ref{sec-gamma2}), we showed the family of elliptic curves $E[2]$ with level structure for $\Gamma(2)$  is equivalent to a universal family of complex tori  $\mathcal{E}_{\Gamma(2)\backslash{\mathbb{H}}}$.
%and in eq.~(\ref{eq:modular-g14}) we realized that 
The periods  are modular forms of weight 1 on  $\mathcal{E}_{\Gamma_1(4)\backslash{\mathbb{H}}}\equiv\left(\mathbb{Z}^2\rtimes\Gamma_1(4)\right)\backslash\left(\mathbb{C}\times{\mathbb{H}}\right)$.
A natural question arises: what is the algebraic counterpart of $\mathcal{E}_{\Gamma_1(4)\backslash{\mathbb{H}}}$?
%Translated another way, given some congruence subgroup, say the Fuchsian triangle group $\Gamma_1(4)$, 
%on which family of elliptic curves  the relevant torsion data of $\Gamma_1(4)$ is realized?
That family of elliptic curves is
\beq
 \label{eq:Et4-def}
 E_{t_4}: Y^2=(X^2-1)(X^2-t_4)\,, 
 \eeq
 where 
$t_4 \in \mathbb{CP}^1\setminus\Sigma$ with $\Sigma= \{0,1,\infty\}$. The omitted  points are those when $E_{t_4}$    degenerates.
The family of periods can be expressed in terms of complete elliptic integrals of the first kind,
where
\begin{equation}
	\begin{split}
  \label{eq:psit4_def}
  \Psi_1(t_4)  =\frac{4 \EK\left(\frac{4\sqrt{t_4}}{(1+\sqrt{t_4})^2}\right)}{1+\sqrt{t_4}} \,,\quad
  \Psi_2(t_4)  =\frac{4 \EK\left(1-\frac{4\sqrt{t_4}}{(1+\sqrt{t_4})^2}\right)}{1+\sqrt{t_4}} 
  \,.
  \end{split}
\end{equation}
The corresponding Picard-Fuchs operator describing a family of elliptic curves is:
\begin{align}
	\label{eq:t4P-F}
	\mathcal{L}:=\partial_{t_4}^2 + \left(\frac{1}{t_4}-\frac{1}{1-t_4}\right)\partial_{t_4} +
	\frac{1}{4(t_4-1)t_4}
\,.
\end{align}
The images of the generators of $\pi_1$ under the  homeomorphism eq.~(\ref{mono-rep}) are given by  
\begin{align}
\rho_{[{\circlearrowleft}_0]}=
\begin{pmatrix}
     1 & 1 \\
    0 & 1
  \end{pmatrix}\,,
  \quad
  \rho_{[{\circlearrowleft}_1]}=
  \begin{pmatrix}
     1 & 0 \\
    -4 & 1
  \end{pmatrix}\,,
  \end{align}
which are precisely the two candidate generators for the free group $\Gamma_1(4)$.
Thus $\rho(\pi_1(X,\cdot))=\langle \rho_{[{\circlearrowleft}_0]}\,, \rho_{[{\circlearrowleft}_1]} \rangle =\Gamma_1(4)$.
The Hauptmodul identifying $\mathbb{CP}^1\setminus\Sigma$ with $\Gamma_1(4)\backslash \mathbb{H}$ is~\cite{maier2008rationally}
\beq
\label{eq:t4-def}
t_4(\tau)=\left(\frac{\theta_3^2(\tau)-\theta_4^2(\tau)}{\theta_3^2(\tau)+\theta_4^2(\tau)}\right)^2\,.
\eeq
Furthermore, the pullback of the period function by $t_4$ is given by
  \beq
  \label{eq:modularg4}
 \Psi_1(\tau) =4\EK(t_4)= \pi (\theta_3^2(q)+\theta_4^2(q))=2\pi \theta^2_3(q^2)
 =2\pi\frac{\eta^{10}(2\tau)}{\eta(\tau)\eta^4(4\tau)}\,,
 \eeq
 indeed, it defines a modular form of weight 1 for $\Gamma_1(4)$, 
and note that dim$(\mathcal{M}_1(\Gamma_1(4)))=1$.

We establish the equivalence between 
$\mathcal{E}_{\Gamma_1(4)\backslash{\mathbb{H}}}$
and the following family of elliptic curves
\begin{align}
\label{eq:univer-g14}
 E_{\tau}[4]:  Y^2=(X^2-1)(X^2-t_4(\tau))
 \,, \quad  \tau \in \Gamma_1(4)\backslash \mathbb{H}\,,
 \end{align}
where $t_4(\tau)$ is the Hauptmodul  for $\Gamma_1(4)$ given by eq.~(\ref{eq:t4-def}).
We define a family of inverse  Abel maps from $\mathcal{E}_{\Gamma_1(4)\backslash{\mathbb{H}}}$ to $ E_{\tau}[4]$,
given by $f_{[4]}$
\beq
(z,\tau) \in \mathcal{E}_{\Gamma_1(4)\backslash{\mathbb{H}}} \stackrel{f_{[4]}}{\longmapsto} \left[X: \frac{1}{\Psi_1(\tau)}\partial{X}/{\partial z}:1\right] \in  E_{\tau}[4]\,,
\eeq
where
 the period function $\Psi_1(\tau)$ 
 is given in eq.~(\ref{eq:modularg4}) and the $X$-coordinate is given by
\beq
\label{eq: modular-para-g14}
2 \frac{\sqrt{t_4(\tau)}+X}{(\sqrt{t_4(\tau)}+1)(1+X)}=\frac{\theta_2^2(\tau)}{\theta_3^2(\tau)}\frac{\theta_4^2(z,\tau)}{\theta_1^2(z,\tau)}\,.
\eeq
 The function $f_{[4]}$ defined in eq.~(\ref{eq: modular-para-g14}) is invariant under the action of $\mathbb{Z}^2\rtimes\Gamma_1(4)$,
\beq
f_{[4]}[z\,,\tau]=f_{[4]}[((m,n),\gamma)\cdot(z\,,\tau)]\,,\quad\forall  (m\,,n) \in  \mathbb{Z}^2\,, \gamma\in\Gamma_1(4)\,.
\eeq
%where $((m,n),\gamma)\cdot(z\,,\tau)=\left( \frac{z+m \tau+n}{c \tau+d}\,,\gamma\cdot\tau \right)$.
Thus, $f_{[4]}$ descends to a well-defined holomorphic map from $\mathcal{E}_{\Gamma_1(4)\backslash{\mathbb{H}}}$ to $E_{\tau}[4]\simeq E_{t_4}$.
The map is bijective, its inverse is given by a family of Abel maps, we say $E_{t_4}$ is an algebraic realization of  the universal curves $\mathcal{E}_{\Gamma_1(4)\backslash{\mathbb{H}}}$.

Here is a short summary: we have identified $(z\,,\tau)$ on $\mathcal{E}_{\Gamma_1(4)\backslash{\mathbb{H}}}=\left(\mathbb{Z}^2\rtimes\Gamma_1(4)\right)\backslash\left(\mathbb{C}\times{\mathbb{H}}\right)$ with  $(X\,,t_4)$ on $E_{t_4}$, 
through a biholomorphic  map $\tilde{g}$ given by the Hauptmodul eq.~(\ref{eq:t4-def}) and the family of Abel maps in eq.~(\ref{eq: modular-para-g14}):
\beq
\label{diagram:commu-high-dimen-2}
\begin{tikzpicture}[scale=2]
    \node (E) at (0,0) {$\mathbb{C}\times{\mathbb{H}}$};
    \node[below=of E] (N) {$\left(\mathbb{Z}^2\rtimes\Gamma_1(4)\right)\backslash\left(\mathbb{C}\times{\mathbb{H}}\right)$};
    \node[right=of N] (M) {$\mathbb{C}\mathbb{P}^2\setminus\{\text{kinematic branches}\}$};
    \draw[->] (E)--(N) node [midway,left] {$\pi$};
    \draw[->,transform canvas={yshift=+1mm}] (N)--(M) node [midway,above] {$\tilde{f}$};
    \draw[->,transform canvas={yshift=-1mm}] (M)--(N) node [midway,below] {$(z, \Psi_2/\Psi_1)$};
    \draw[->] (E)--(M) node [midway,above] {$f$};
     \node[left=of N] (Q) {$ E_{t_4}: Y^2=(X^2-1)(X^2-t_4)$};
      \draw[->] (E)--(Q) node [midway,above] {$g$};
       \draw[->,transform canvas={yshift=+1mm}] (N)--(Q) node [midway,above] {$\tilde{g}$};
        \draw[->,transform canvas={yshift=-1mm}] (Q)--(N) node [midway,below] {$\tilde{g}^{-1}$};
\end{tikzpicture}
\eeq
The effect of such an isomorphism is twofold: First, we can reparametrize the Mandelstam variables for Bhabha using canonical coordinates $\left([x:Y:1]\,,t_4\right)\in E_{t_4}$
\beq
\label{eq:to-uni-curve}
\mathrm{s}=2\frac{(1-x)(1+t_4)}{t_4-x}\,,\quad \mathrm{t}=4 \frac{t_4 (1-x^2)}{x^2-t_4^2}\,.
\eeq
Unlike the parametrization in eq.~(\ref{eq:modular-para-cp2}), this transformation is fully algebraic. We can clearly see all the branches of the full set of the symbol letters,
e.g.\,, 
we can compute the period of Bhabha scattering 
% \beq
%\pi \sqrt{-t}\sqrt{4-t} \Psi_{\text{bhabha}}(x,t_4)=2\frac{\sqrt{t_4^2-x^2}\EK(\lambda)}{\sqrt{1+t_4}   \sqrt{1-x^2}   }\,,
% \eeq
  \beq
   \Psi_{\mathrm{bhabha}}(x,t_4)=\frac{2(x^2-t_4)}{- Y  }\EK(t_4)\,,
% \Psi_{\mathrm{bhabha}}(x,t_4)=\frac{2\sqrt{t_4-x^2}\EK(\lambda)}{(1+\sqrt{t_4})   \sqrt{1-x^2}   }\,,
 \eeq
where $ \EK(t_4)$ is the unique modular form of weight one for $\Gamma_1(4)$ in eq.~(\ref{eq:modularg4}),
and   $Y$ is the $Y$-coordinate of the universal family of elliptic curves $E_{t_4}$ given by eq.~(\ref{eq:Et4-def}).
We can also compute again for the pullback of the symbol letters given in section~\ref{pullback-bhabha}:
\begin{align}
  \omega_z\mapsto  \frac{\pi}{2 \EK (t_4)} \left(\frac{\rd x}{Y}+\mathcal{F}(x,t_4) \rd t_4 \right)\,,
\quad
    \omega_{\tau}\mapsto
       \frac{\pi^2}{8 \EK^2(t_4)}
   \frac{1}{t_4 (1-t_4)}
     \rd t_4\,,
 \end{align}
\begin{align}
    \omega_{41}\mapsto& \left[
    2\frac{1-x^2}{(1-t_4)(t_4-x^2)}
    +
  8
   (1-t_4)
    t_4
    \mathcal{F}^2(x,t_4)
    \right] \rd t_4
    +
   16 (1-t_4) t_4\mathcal{F}(x,t_4)
 \frac{\rd x}{Y}
    \,,
 \end{align}
where $\mathcal{F}(x,t_4)$ is the $\tau$-derivative of the Abel maps
\beq
 \label{eq:AbelG4}
\mathcal{F}(x,t_4)=\EK(t_4)\times\partial_{t_4}{\left[\frac{1}{\EK(t_4)}F\left(x|t_4\right)\right]}\,,
\quad
 F\left(x|t_4\right)\equiv \int_{-1}^{x} \frac{d X}{\sqrt{(X^2-1)(X^2-t_4)}}\,.
 \eeq
 $ \mathcal{F}(x,t_4)$ is related to  $Z_4(x,t_4)$ for the eMPLs~\cite{Broedel:2017kkb}
\beq
 \mathcal{F}(x,t_4)=\frac{1}{4 t_4}\frac{Z_4(x,t_4)}{\sqrt{t_4}-1}-\frac{x Y}{2 t_4(t_4-1)(x^2-t_4)}\,.
 \eeq
On the other hand, we can have an algebraic realization of Kronecker's differential forms, e.g.\,,
\begin{align}
2\pi i\omega^{\text{Kro}}_{3}(2z,\tau)
\xmapsto{~\tilde{g}^{{-1}*}}&\,
\rd x
\EK(t_4)
 \left[
 \frac{16t_4^2 (1-t_4)^2}{Y}\mathcal{F}^2(x,t_4)
+\frac{8 t_4 (1-t_4)^2 x}{Y^2}\mathcal{F}(x,t_4)
-\frac{2}{3}\frac{1+t_4}{Y}
 \right]
  \nonumber\\
 + &\rd t_4
\EK(t_4)
 \left[
 \frac{16}{3}(1-t_4)^2t_4^2 \mathcal{F}^3(x,t_4)
 + \frac{2}{3} (1+t_4) \mathcal{F}(x,t_4)
 +\frac{2}{3}\frac{x(x^2-1)}{Y(x^2-t_4)}
  \right]
 \,.
\end{align}
%For more identities, please refer to the main paper, there
%we build a translation table for $\Gamma_1(4)$ up to modular weight 4.

  %%%%%%%%%%%%%%%%%%%%%%%%%%%%%%%%%%%%%%%
\section{Conclusion}
  %%%%%%%%%%%%%%%%%%%%%%%%%%%%%%%%%%%%%%%
We computed Bhabha scattering at two loops--an amplitude beyond genus $0$ in QED. 
We revealed underlying connections between this amplitude and the arithmetic geometry of elliptic curves. 
We gave unified descriptions for several sectors of Bhabha scattering and planar top quark production 
through canonical coordinates on the moduli space $\mathscr{M}_{1;2}[4]$.
Finally, we established correspondence between the Kronecker's differential forms and letters of eMPLs.

\section*{Acknowledgements}
This work was co-funded by
the European Union through the ERC Consolidator Grant LoCoMotive 101043686. Views and opinions expressed are
however those of the author(s) only and do not necessarily reflect those of the European
Union or the European Research Council. Neither the European Union nor the granting
authority can be held responsible for them.

{\footnotesize
\bibliography{Bhabha}
\bibliographystyle{h-physrev}
}

%\begin{thebibliography}{99}
%
%%\cite{Zhu:2020ftr}
%\bibitem{Zhu:2020ftr}
%Y.~J.~Zhu,
%%``Double soft current at one-loop in QCD,''
%[arXiv:2009.08919 [hep-ph]].
%%18 citations counted in INSPIRE as of 05 Sep 2023
%
%\end{thebibliography}

\end{document}